\documentclass[a4paper, 10pt]{article}

\usepackage{style/osameet2}

\usepackage[acronym,nomain]{glossaries}
\usepackage{amsmath,amssymb}
\usepackage{textgreek}
\usepackage{xfrac}
\usepackage{subfig}    

\usepackage[pdftex,colorlinks=true,bookmarks=false,citecolor=blue,urlcolor=blue]{hyperref} 

\newacronym{CUT}{CUT}{channel under test}

\newacronym{KK}{KK}{Kramers-Kronig}
\newacronym{CSPR}{CSPR}{carrier-to-signal power ratio}
\newacronym{KKRX}{KKRx}{Kramers-Kronig Receiver}
\newacronym{SSBI}{SSBI}{signal-signal beat interference}

\newacronym{GS}{GS}{geometric shaping}
\newacronym{PS}{PS}{probabilistic shaping}
\newacronym{DSP}{DSP}{digital signal processing}
\newacronym{MIMO}{MIMO}{multiple-input multiple-output}
\newacronym{TDE}{TDE}{time domain equalizer}
\newacronym{FDE}{FDE}{frequency domain equalizer}
\newacronym{LMS}{LMS}{least means square}
\newacronym{DDLMS}{DD-LMS}{decision directed least means square}
\newacronym{FFE}{FFE}{feed-forward equalizer}
\newacronym{FBE}{FBE}{feedback equalizer}
\newacronym{BPS}{BPS}{blind phase search}

\newacronym{SMF}{SMF}{single-mode fiber}
\newacronym[plural=SSMFs]{SSMF}{SSMF}{standard single-mode fiber}
\newacronym[plural=FMFs]{FMF}{FMF}{few-mode fiber}
\newacronym{FMF12}{FMF12}{12 mode FMF}
\newacronym{MMF}{MMF}{multi-mode fiber}
\newacronym{SI}{SI}{step index}
\newacronym{GI}{GI}{graded index}
\newacronym{DCF}{DCF}{dispersion compensated fiber}

\newacronym{SDM}{SDM}{space division multiplexing}
\newacronym{MDM}{MDM}{mode division multiplexed}
\newacronym{WDM}{WDM}{wavelength division multiplexing}
\newacronym{DWDM}{DWDM}{dense wavelength division multiplexing}
\newacronym{LP}{LP}{linear polarized}
\newacronym[plural=MMUXs,firstplural=mode multiplexers]{MMUX}{MMUX}{mode multiplexer}
\newacronym{PL}{PL}{photonic lantern}
\newacronym{3DWG}{3DWG}{3D-waveguide}
\newacronym{MDL}{MDL}{mode dependent loss}
\newacronym{DGD}{DGD}{differential group delay}
\newacronym{DMGD}{DMGD}{differential mode group delay}
\newacronym{QSM}{QSM}{quasi-single-mode}
\newacronym{GIMMF}{GI-MMF}{graded-index multi-mode fiber}

\newacronym{SSB}{SSB}{single side band}
\newacronym{QPSK}{QPSK}{quadrature phase shift keying}
\newacronym{QAM}{QAM}{quadrature amplitude modulation}
\newacronym{RRC}{RRC}{root-raised-cosine}
\newacronym{4D-64PRS}{4D-64PRS}{
four-dimensional 64-ary polarization-ring-switching}
\newacronym{DP}{DP}{dual-polarization}
\newacronym[\glslongpluralkey=states-of-polarization]{SOP}{SOP}{state-of-polarization}
\newacronym{PM}{PM}{polarization-multiplexed}

\newacronym{ECL}{ECL}{external cavity laser}
\newacronym{CW}{CW}{continuous wave}
\newacronym[plural=DFBs]{DFB}{DFB}{distributed feedback laser}

\newacronym[plural=DACs]{DAC}{DAC}{digital-to-analog converter}
\newacronym{ADC}{ADC}{analog-to-digital converter}
\newacronym{PRBS}{PRBS}{pseudo-random bit sequence}
\newacronym{LO}{LO}{local oscillator}
\newacronym{EDFA}{EDFA}{erbium-doped fiber amplifier}
\newacronym{MZM}{MZM}{Mach-Zehnder modulator}
\newacronym{DP-MZM}{DP-MZM}{dual-polarization Mach-Zehnder modulator}
\newacronym{ChUT}{ChUT}{channel under test}
\newacronym{WSS}{WSS}{wavelength selective switch}
\newacronym[plural=VOAs]{VOA}{VOA}{variable optical attenuator}
\newacronym[plural=PDCRXs]{PDCRX}{PDCRX}{polarization diverse coherent receiver}
\newacronym{DSO}{DSO}{digital storage oscilloscope}
\newacronym{ASE}{ASE}{amplified spontaneous emission}
\newacronym{PBS}{PBS}{polarization beam splitter}
\newacronym{PD}{PD}{photodiode}
\newacronym{AOM}{AOM}{acousto-optical modulator}
\newacronym{BPD}{BPD}{balanced photo-diode}
\newacronym{OMFT}{OMFT}{optical-multi-format transmitter}
\newacronym{DPIQ}{DP-IQM}{dual-polarization IQ-modulator}
\newacronym{ABC}{ABC}{automatic bias controller}
\newacronym{OTF}{OTF}{optical tunable filter}
\newacronym{LSPS}{LSPS}{loop-synchronous polarization scrambler}
\newacronym{OSA}{OSA}{optical spectrum analyzer}

\newacronym{OSNR}{OSNR}{optical signal to noise ratio}
\newacronym{BER}{BER}{bit error rate}
\newacronym{IL}{IL}{insertion loss}

\newacronym{SDFEC}{SD-FEC}{soft-decision forward error correction}
\newacronym{HDFEC}{HD-FEC}{hard-decision forward error correction}
\newacronym{FEC}{FEC}{forward error correction}
\newacronym{LDPC}{LDPC}{low-density parity-check code}

\newacronym{AIR}{AIR}{achievable information rate}
\newacronym{AR}{AR}{achievable rates}
\newacronym{MI}{MI}{mutual information}
\newacronym{GMI}{GMI}{generalized mutual information}
\newacronym{BICM}{BICM}{bit-interleaved coded modulation}

\newacronym{OVNA}{OVNA}{optical vector network analyzer}
\newacronym{NIR}{NIR}{near infrared}
\newacronym{CD}{CD}{chromatic dispersion}
\newacronym{OTDR}{OTDR}{optical time domain reflectometry}
\newacronym{OFDR}{OFDR}{optical frequency domain reflectometry}
\newacronym{GPU}{GPU}{graphics processing unit}
\newacronym{SVD}{SVD}{singular value decomposition}
\newacronym{WGN}{WGN}{white Gaussian noise}
\newacronym{AWGN}{AWGN}{additive white Gaussian noise}
\newacronym{PDL}{PDL}{polarization dependent loss}
\newacronym{SPS}{sps}{samples-per-symbol}
\newacronym{SE}{SE}{spectral efficiency}

\begin{document}



\title{11,700 km Transmission at 4.8 bit/4D-sym via  Four-dimensional Geometrically-shaped Polarization-Ring-Switching Modulation}

 \vspace{-4mm}
\author{
  Sjoerd~van~der~Heide\textsuperscript{(1)},
  Bin~Chen\textsuperscript{(1)(3)},
  Menno van den Hout\textsuperscript{(1)},
  Gabriele Liga\textsuperscript{(1)},\\
  Ton Koonen\textsuperscript{(1)},
  Hartmut~Hafermann\textsuperscript{(2)},
  Alex~Alvarado\textsuperscript{(1)},
  and~Chigo~Okonkwo\textsuperscript{(1)}}
\address{
  \textsuperscript{(1)}Department of Electrical Engineering, Eindhoven University of Technology, 
  The Netherlands.\\  
  \textsuperscript{(2)}Mathematical and Algorithmic Sciences Lab, Paris Research Center, Huawei Technologies France SASU, 
  France.\\  
  \textsuperscript{(3)}School of Computer Science and Information Engineering, Hefei University of Technology, China
 \vspace{-2mm}
  \email{s.p.v.d.heide@tue.nl}}
\vspace{-7mm}

 \begin{abstract}
Using a novel geometrically-shaped four-dimensional modulation format, we transmitted $11\times200$~Gbit/s DWDM  at 4.8 bit/4D-sym over 7,925~km and 11,700~km using EDFA-only and hybrid amplification, respectively. A reach increase of 16\%  is achieved over PM-8QAM.
\end{abstract}

\vspace{-2mm}

\vspace{-2mm}
\section{Introduction}
The non-linear Shannon limit for single-mode transmission systems remains despite the strong growth in traffic. 
In recent years, signal shaping via \gls{PS} or \gls{GS} have been demonstrated to close the gap to the non-linear Shannon limit by employing non-uniform probability or non-equidistant constellation points whilst maximising \gls{AIR}, respectively. Both techniques have been extensively demonstrated with \gls{PS} shown to require complex coding techniques, whilst \gls{GS} only requires straightforward modifications of the mapper and demapper.

Multidimensional constant modulus modulation formats have been shown to minimize the non-linear interference noise by minimizing the signal power variations \cite{ReimerOFC2016,Kojima2017JLT}. Recently, the \gls{4D-64PRS} format was introduced in \cite{BinChenArxiv2019}, where the 4D coordinates and labeling were jointly optimized. Numerical results in \cite{BinChenArxiv2019} show that \gls{4D-64PRS} outperforms other modulation formats with \gls{SE} of 6 bit/4D-sym, including \gls{PM}-8QAM, 4D-64SP-12QAM \cite{NakamuraECOC2015} and 4D-2A8PSK \cite{Kojima2017JLT}. Results were presented in both linear and non-linear channel for a \gls{BICM} system \cite{BinChenArxiv2019}. \gls{4D-64PRS} was reported to yield over 0.8 dB better sensitivity than PM-8QAM by maximizing \gls{GMI} \cite{BinChenArxiv2019}.



In this work, we  experimentally show a  16\% reach increase over \gls{PM}-8QAM, which is considered a viable candidate for beyond 200G per optical carrier ultra long-haul transmission \cite{OIF400G,ZhangOFC2014,ZhangJLT2018}.  Using a novel four-dimensional 4D-64PRS modulation format, both EDFA-only and hybrid EDFA plus Raman amplified long-haul  transmission scenarios are evaluated. 
4D-64PRS is shown to outperform both \gls{PM}-8QAM and 6b4D-2A8PSK in an 11-channel \gls{DWDM}, net 200 Gbit/s/channel \gls{SSMF} recirculating loop transmission experiment. It is shown that pre-FEC reach gains are preserved for post-FEC.

\vspace{-2mm}
\section{Experimental transmission setup}
\begin{figure}[!t]
  \centering
  \includegraphics[width=\textwidth]{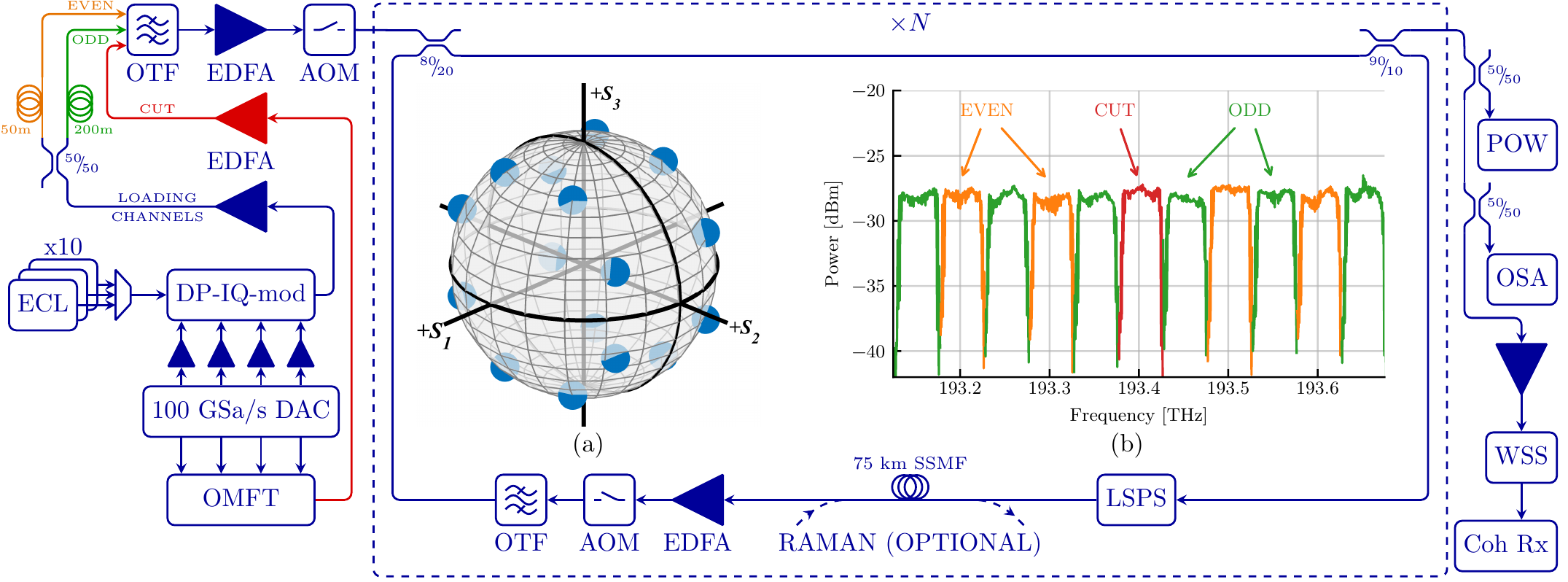}
  \caption{Experimental setup. Inset \textbf{(a)} shows a Stokes-space representation of the \gls{4D-64PRS} modulation format, indicating 64 constant-modulus points in 16 distinct \glspl{SOP}. Inset \textbf{(b)}  shows the received spectrum after 106 circulations (8,000~km) of EDFA-only amplification. Note that the \gls{CUT} is depicted in the center position but is tested in all 11 positions in the experiment.}
  \label{fig:setup}
  \vspace{-6mm}
\end{figure}

\begin{figure}[!b]
\vspace{-6mm}
  \centering
  \subfloat[]{\includegraphics[width=.4\textwidth]{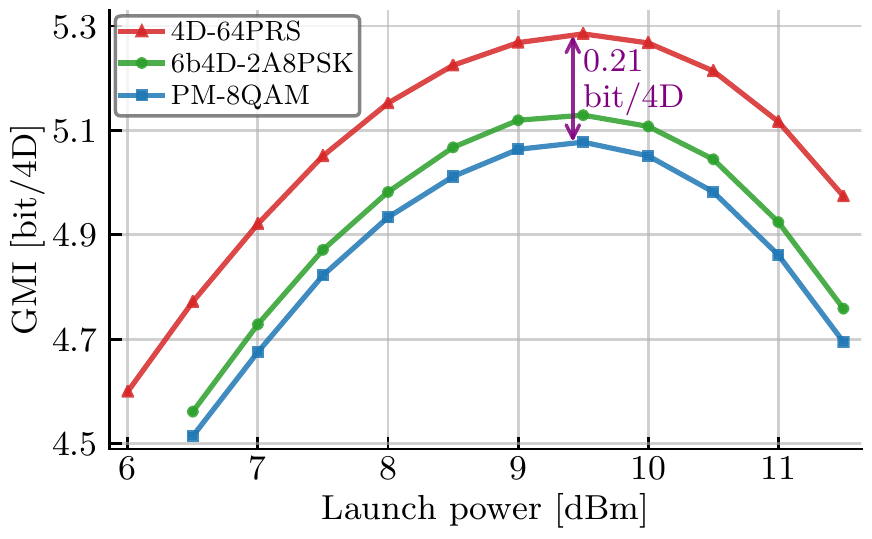}\label{fig:edfalp}}
  \subfloat[]{\includegraphics[width=.2\textwidth]{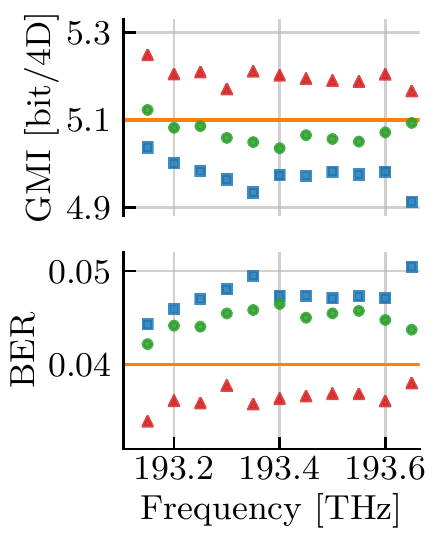}\label{fig:edfach}}  
  \subfloat[]{\includegraphics[width=.4\textwidth]{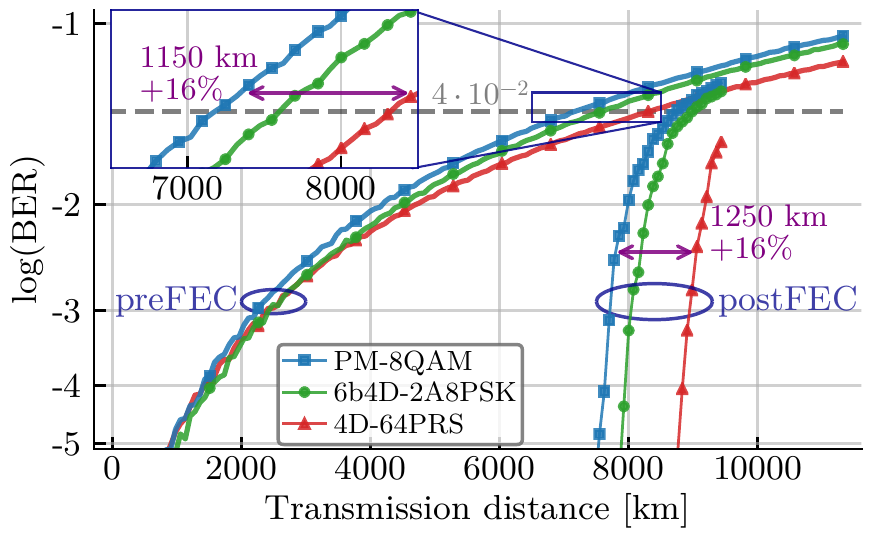}\label{fig:edfadistance}}
  \vspace{-2mm}
\caption{Experimental results using EDFA-only amplification.
\textbf{(a)} Average GMI per channel versus total launch power after 7,550~km.
\textbf{(b)} Per-channel performance versus BERs (top) and GMIs (bottom) measured for all 11 channels individually after 7,925~km showing BERs below the FEC threshold $4\cdot10^{-2}$\cite{Kojima2017JLT} (top) and GMIs above 5.1~bit/4D\cite{scfec} (bottom) for all 11 channels for 4D-64PRS.
\textbf{(c)} BER versus transmission distance for the center channel at launch power of 9.5~dBm. 
}

\label{fig:edfa}
\vspace{-6mm}
\end{figure}

Fig.~\ref{fig:setup} depicts the experimental transmission setup. 
The transmitted signal is  modulated using eithr PM-8QAM, 6b4D-2A8PSK \cite{Kojima2017JLT}, or \gls{4D-64PRS} \cite{BinChenArxiv2019}.
Fig.~\ref{fig:setup} inset (a) shows the Stokes representation of the \gls{4D-64PRS} format, which exhibits 16 distinct \glspl{SOP} and consists of 64 4D-symbols. Each distinct SOP represents four 4D-symbols on the Poincar\'{e} sphere indicating constant modulus in 4D. 
Sequences containing 2\textsuperscript{16} symbols are shaped using a \gls{RRC} filter with 1\% roll-off at 41.79~GBd, generated offline and uploaded to a 100-GSa/s \gls{DAC}. The positive ends of the differential \gls{DAC} outputs are connected to the \gls{OMFT} which consists of an \gls{ECL}, a \gls{DPIQ}, an \gls{ABC} and RF-amplifiers. The \gls{CUT}, which can be defined at any of the 11 tested C-band channels, is modulated by the \gls{OMFT} and subsequently amplified. The loading channels are provided by the negative outputs of the \gls{DAC} and modulated on the tones provided by 10 \glspl{ECL} using a \gls{DPIQ}. These loading channels are amplified, split into even and odd, decorrelated by 10,200 (50~m) and 40,800 symbols (200~m), and multiplexed together with the \gls{CUT} on a 50-GHz grid using an \gls{OTF}. Bandwidth limitations due to transmitter electronics are compensated for partly through digital filters and partly using the \gls{OTF} as proposed in \cite{Lin2018}. 

The 11-channel 50-GHz-spaced \gls{DWDM} signal is amplified and through an \gls{AOM} enters the recirculating loop which consists of a \gls{LSPS}, a 75-km span of \gls{SSMF}, an \gls{EDFA}, an \gls{AOM} and an \gls{OTF} used for gain flattening. Fig.~\ref{fig:setup} inset (b) shows the optical spectrum after 106 circulations, which corresponds to 8,000~km of transmission using only \gls{EDFA}-amplification. Optionally, a hybrid amplification scheme can be used by adding a 750~mW 1480~nm Raman pump in a backward configuration. Part of the output of the recirculating loop is measured by a power meter, part is analyzed using an \gls{OSA} and the rest is amplified, filtered by a \gls{WSS} and digitized by a coherent receiver consisting of a \gls{LO}, a 90-degree hybrid, four balanced photo-diodes and an 80-GSa/s \gls{ADC}. Offline \gls{DSP} includes front-end correction, frequency-offset compensation, chromatic dispersion compensation, \gls{MIMO} equalization with in-loop \gls{BPS}, error counting and \gls{GMI} evaluation.

\vspace{-2mm}
\section{Results}
\begin{figure}[!t]
  \centering
  \subfloat[]{\includegraphics[width=.4\textwidth]{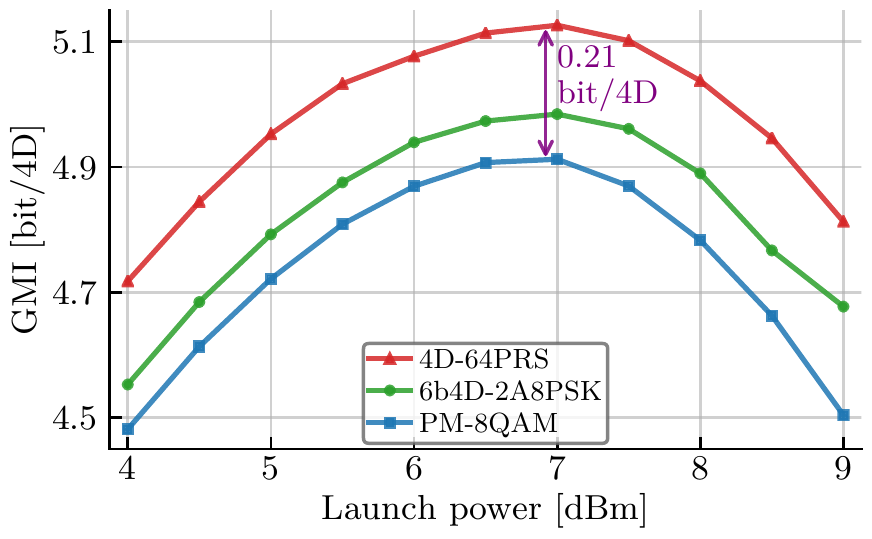}\label{fig:hybridlp}}
  \subfloat[]{\includegraphics[width=.2\textwidth]{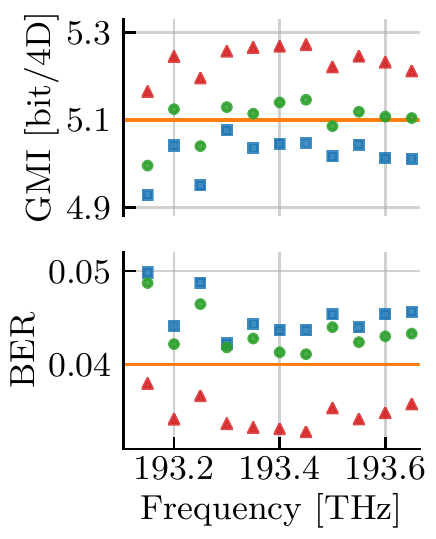}\label{fig:hybridch}}  
  \subfloat[]{\includegraphics[width=.4\textwidth]{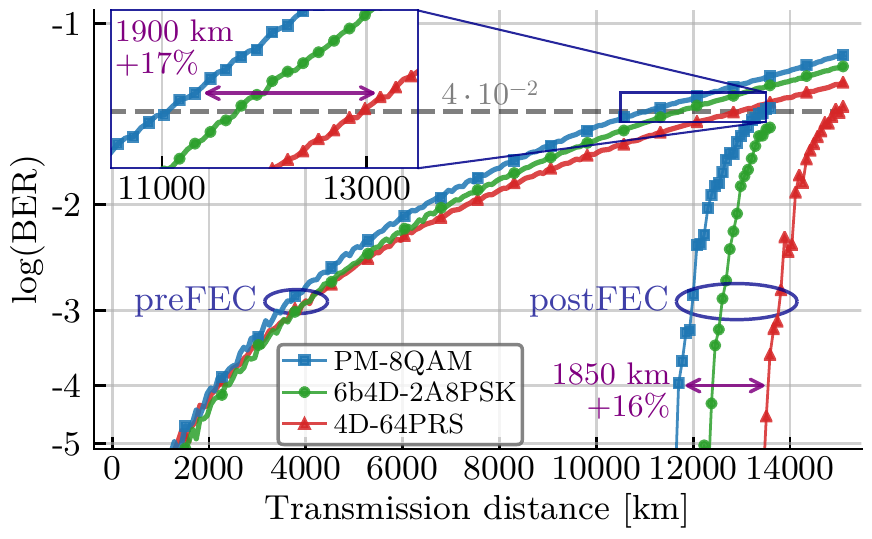}\label{fig:hybriddistance}}
    \vspace{-2mm}
\caption{Experimental results using a hybrid of EDFA and Raman amplification.
\textbf{(a)} Average GMI per channel versus total launch power after 12,530~km.
\textbf{(b)} Per-channel performance versus BERs (top) and GMIs (bottom) measured for all 11 channels individually after 11,700~km showing BERs below the FEC threshold $4\cdot10^{-2}$\cite{Kojima2017JLT} (top) and GMIs above 5.1~bit/4D\cite{scfec}.
\textbf{(c)} BER versus transmission distance for the center channel at launch power of 6.5~dBm.}

\label{fig:hybrid}
\vspace{-6mm}
\end{figure}

Transmission scenarios using EDFA-only amplification (Fig. \ref{fig:edfa}) and a hybrid of EDFA and Raman (Fig. \ref{fig:hybrid}) are evaluated. Fig. \ref{fig:edfalp} shows that a total launch power of 9.5~dBm maximizes the average \gls{GMI} per channel resulting in a 0.21~bit/4D increase for 4D-64PRS with respect to PM-8-QAM. At this optimal launch power, we demonstrate transmission below the pre-FEC-threshold of $4\cdot10^{-2}$ and above the \gls{GMI} threshold of 5.1~bit/4D, thus enabling error-free 7,925~km transmission of net 200 Gbit/s per channel after 25.5\% overhead and for all 11 channels. The GMI threshold 5.1~bit/4D (0.85 NGMI) is based on a spatially-coupled type LDPC code\cite{scfec} and the corresponding BER threshold of $4\cdot10^{-2}$ is derived in \cite{Kojima2017JLT}.

A reach increase of  16\% (+1,150~km), for 4D-64PRS compared to \gls{PM}-8QAM is shown for a pre-FEC \gls{BER} of $4\cdot10^{-2}$ in Fig. \ref{fig:edfadistance}.
Moreover, the 16\% reach increase is preserved for the post-FEC gain using an off-the-shelf DVB-S2 \gls{LDPC} with 25\% overhead and code length $n=64800$.
In addition to the gains shown versus \gls{PM}-8QAM, 4D-64PRS is also shown to outperform 6b4D-2A8PSK. \gls{BER} and \gls{GMI} calculations were done using over 42 million bits for all plotted data points. Note that 63 million bits were used for \gls{LDPC} decoding.

For the hybrid amplification scenario, the relative gains are similar to the EDFA-only case as shown in Fig. \ref{fig:hybrid}. A \gls{GMI}-increase of 0.21~bit/4D at 7~dBm launch power is observed in Fig.~\ref{fig:hybridlp}. Further measurements were carried out at 6.5~dBm total launch power, near the optimal launch power. Fig. \ref{fig:hybridch} shows all channels were able to transmit over 11,700~km successfully. Fig. \ref{fig:hybriddistance} shows a pre-FEC reach increase of 17\% and a post-FEC reach increase of 16\%, which is consistent with the EDFA-only amplification scenario.

\vspace{-2mm}
\section{Conclusions}
The novel geometrically-shaped 4D-64PRS modulation format is experimentally compared to other notable 6 bit/4D modulation formats such as \gls{PM}-8QAM and 6b4D-2A8PSK, with 4D-64PRS outperforming both of them. Experimental results show a reach increase of 16\% at 7,925~km for \gls{SSMF} and EDFA-only amplification and 11,700~km of using hybrid of both EDFA and Raman amplification. \gls{LDPC} decoding is performed on the experimental data to demonstrate a post-FEC 16\% reach increase at a net \gls{SE} of 4.8~bit/4D. 

\vspace{-2mm}
\noindent\small\emph{\\Partial funding from the Dutch NWO Gravitation Program on Research Center for Integrated Nanophotonics (Grant Number 024.002.033). Research supported by Huawei France through the NLCAP project. The work of A. Alvarado is supported by the Netherlands Organisation for Scientific Research (NWO) via the VIDI Grant ICONIC (project number 15685). Fraunhofer HHI is acknowledged for providing their Optical-Multi-Format Transmitter.}

\vspace{-2mm}
\bibliographystyle{style/osajnl}
\bibliography{ref.bib}

\end{document}